\documentclass[twocolumn,showpacs,superscriptaddress,amsmath,amssymb,aps,pre]{revtex4}
\usepackage{dcolumn}% Align table columns on decimal point
\usepackage{bm}% bold math
\usepackage[dvips]{graphicx}
\usepackage{wrapfig,subfigure}
\usepackage{amssymb}
\usepackage{color}
\usepackage{ulem}

\newcommand{\qS}{q_{\cal S}}

\newcommand{\la}{\lambda_a}
\newcommand{\lme}{\lambda_{\rm esc}}
\newcommand{\lS}{\lambda_{\cal S}}

\begin{document}

\title{Poisson-noise induced escape from a metastable state}

\author{M. I. Dykman}
%\email{dykman@pa.msu.edu}
\affiliation{Department of Physics and Astronomy, Michigan State University, East Lansing, MI 48824}
\date{\today}

\begin{abstract}
We provide a complete solution of the problems of the probability distribution and the escape rate in Poisson-noise driven systems. It includes both the exponents and the prefactors. The analysis refers to an overdamped particle in a potential well. The results apply for an arbitrary average rate of noise pulses, from slow pulse rates, where the noise acts on the system as strongly non-Gaussian, to high pulse rates,  where the noise acts as effectively Gaussian.
\end{abstract}

\pacs{05.40.Ca, 72.70.+m, 05.70.Ln, 05.40.Jc }

\maketitle

\section{Introduction}
\begin{figure}[h]
\begin{center}
\includegraphics[width=2.0in]{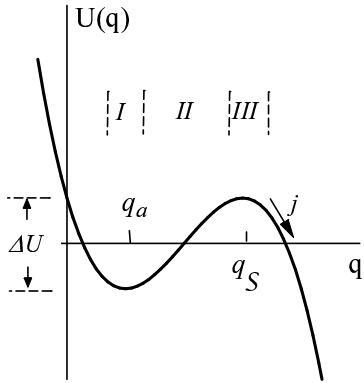}
\end{center}
\caption{A sketch of a metastable potential. The barrier height is $\Delta U=U(\qS)-U(q_a)$. In the time range $t_r\ll t\ll W^{-1}$ the probability current from the metastable state $j$ is independent of time and the escape rate $W=j$ \cite{Kramers1940}. The probability distribution for Poisson noise is found by matching the asymptotic solutions in regions $I, II, III$. }
\label{fig:metastable_potential}
\end{figure}

Escape from a metastable state underlies a broad range of phenomena, chemical reactions, diffusion in solids, and population extinction being examples. A classical theory of escape was first developed by Kramers \cite{Kramers1940}. For a Brownian particle in a potential well, see Fig.~\ref{fig:metastable_potential}, he obtained the escape rate $W$ in the form $W=\lme\exp(-\Delta U/k_BT)$ and found the prefactor $\lme$ in a broad parameter range. The understanding of the prefactor is important not only for completeness of a theory of escape, but also for interpreting the experiment, cf. Refs.~\cite{Devoret1987,Turlot1998,McCann1999}. Therefore much effort has been put into extending the Kramers theory, see Refs.~\cite{Landauer1989,Buttiker1989,Melnikov1991} for a review. Most of the obtained results refer to generalizations of the Kramers model to other types of Brownian motion, for example, to Brownian motion in a three-dimensional potential \cite{Landauer1961} or a periodically modulated potential \cite{Larkin1986,Ivlev1986,Linkwitz1991,Smelyanskiy1999,Lehmann2000a,Maier2001a,Dykman2005a}.

In the last few years escape from a metastable state has attracted much interest as a means of detecting non-Gaussian noise and potentially determining its statistics \cite{Tobiska2004,Pekola2004,Ankerhold2007a,Sukhorukov2007,Timofeev2007,Billings2008,Grabert2008,LeMasne2009,Zou2010}. The proposed noise detectors are continuous systems; in the experiment there have been used Josephson junctions \cite{Timofeev2007,LeMasne2009} and nanomechanical resonators \cite{Zou2010}. It is important therefore to have a full theory of escape induced by non-Gaussian noise, which will include both the exponent and the prefactor in the escape rate.

In this paper we study the probability distribution and escape induced by Poisson noise. Such noise is often encountered in photon statistics and in the statistics of current through tunnel junctions. The noise can be far from the Gaussian limit, which happens if the noise pulses are infrequent, with the rate of the order of the reciprocal relaxation time of the system in the absence of noise $t_r^{-1}$ (a more precise condition is specified later). This parameter range was studied in the experiment \cite{Zou2010} in particular. In the opposite limit of frequent small pulses the noise becomes close to Gaussian. This range is often of interest for experiments with Josephson junctions.

The random motion of the system is very different depending on whether the system has time to relax between the pulses or they are too close in time to change the state between them. Respectively, the probability distributions of the system have very different shapes and the expressions for the escape rate have very different structures. At the same time, in the both cases the noise has the same statistics. One might expect therefore that the noise-induced fluctuations can be studied within a single approach that would apply for an arbitrary pulse rate. Developing such an approach and demonstrating how the fluctuations change from the well-understood Gaussian limit to the opposite limit of well-separated pulses, along with the escape problem, provide the major motivation for this paper.

We model the system by a particle in a potential well, with the local minimum of the potential corresponding to the metastable state in the absence of noise, see Fig.~\ref{fig:metastable_potential}. Escape occurs if a sufficiently large outburst of noise drives the particle over the potential barrier. We further assume that the particle is overdamped, it has no inertia. For infrequent noise pulses the probability distribution is singular in this case \cite{Billings2008,Baule2009}. As a consequence, the Kramers approach cannot be applied and a different technique has to be used.

Our analysis is based on the kinetic equation. We develop an asymptotic method of solving this equation in the case where the Poisson noise is weak on average, so that escape is a rare event, the escape rate $W \ll t_r^{-1}$. This technique applies for an arbitrary relation between the two parameters that characterize a Poisson noise, the appropriately scaled noise intensity and pulse rate. We find the probability distribution near the minimum of the potential well (point $q_a$ in Fig.~\ref{fig:metastable_potential}) and near the top of the potential barrier (point $\qS$) as integral transforms of different types. The obtained expressions are then matched to the distribution in the intermediate range, which is found using the WKB-type approximation. This gives the full probability distribution, and both the exponent and the prefactor in $W$.

\section{The model}

The dynamics of an overdamped particle, with coordinate $q$, driven by a Poisson noise $f_P(t)$ is described by the Langevin equation
\begin{eqnarray}
\label{eq:Langevin}
&&\dot q=-U'(q)+f_P(t)-\langle f_P(t)\rangle, \\
&&f_P(t)=g\sum_n\delta (t-t_n).\nonumber
\end{eqnarray}
Here, $U(q)$ is a metastable potential. In the absence of noise, the particle has a stable state at the local potential minimum $q_a$ and an unstable stationary state at the local maximum $\qS$, see Fig.~\ref{fig:metastable_potential}. The characteristic relaxation time is $t_r= \la^{-1}$, where $\la=U''(q_a)$. An important class of systems described by an overdamped one-dimensional particle in a metastable potential are systems near bifurcation points. Their dynamics display model-independent features, and their noise-induced switching has found applications in various areas of physics, see Ref.~\cite{Vijay2009} for a recent review. In this case $q$ is the slow variable, $q_a$ is the position of the attractor along the $q$-axis, and $\qS$ is the position of the saddle point.

Poisson noise $f_P(t)$ is chosen in Eq.~(\ref{eq:Langevin}) to be of the simplest type, a sequence of unipolar pulses of area $|g|$ that occur at independent instants $t_n$. We assume that the sign of $g$ is such that the noise pushes the system from $q_a$ to $\qS$, i.e., $(\qS-q_a)/g>0$. The parameter $g$ is small, which means that many pulses are required for pushing the particle over the barrier,
\[(\qS-q_a)/g \gg 1.\]
We note that $(\qS-q_a)/g$ is not necessarily the largest parameter of the theory, i.e., the scaled area of the noise pulses, even though it is small, is not necessarily the smallest parameter.

The noise-driven dynamics very strongly depend on the average rate $\nu$ at which the noise pulses are repeated. If $\nu t_r$ is not large,  the discreteness of the pulses is of primary importance, the state of the system changes between the pulses when they come at the average rate. On the other hand, in the limit of high mean pulse rate, $\nu t_r\gg 1$, the noise is effectively a white Gaussian noise with intensity $D=\nu g^2/2$.

For a high pulse rate, $\langle f_P(t)\rangle = \nu g$ does not have to be small, it can exceed the characteristic dynamical force $|\qS-q_a|/t_r$ even for small $|g|$. On physical grounds, we incorporate the average bias from the noise $-q\langle f_P(t)\rangle$ into the potential $U(q)$ and study fluctuations induced by the deviations from the mean, i.e., by a zero-mean random force $f_P(t)-\langle f_P\rangle$.

There is a similarity between Poisson-noise driven dynamical systems and reaction systems. The latter are characterized by a large but finite number of elementary units, for example, molecules in a stirred chemical reactor or individuals in a population. The dynamics is controlled by reactions between them. The elementary reactions are short events, which are uncorrelated with each other and are characterized by rates; these events present a Poisson process. A system may have a metastable state, which is approached as a result of the most probable reaction sequence. However, an unlikely yet possible reaction sequence can drive the system far away from this state, leading to switching to a different stable state. For one-species reaction systems, the prefactor in the escape rate can be studied using the Kramers technique \cite{Dykman1995c,Escudero2009} (the prefactor in the probability distribution was missed in Ref.~\cite{Dykman1995c}).

As in reaction systems, in our system the noise leads to a finite increment of the system coordinate. However, our system is continuous, and in addition it is subject to a regular force $-U'(q)$. This leads to a qualitative distinction of the escape problem from that in reaction systems and to the aforementioned inapplicability of the Kramers technique.

We will study escape using the kinetic equation for the probability distribution of the system $\rho\equiv \rho(q,t)$. It has a form \cite{vanKampen_book}
\begin{eqnarray}
\label{eq:Fokker_Planck}
\partial_t\rho(q,t) &=&
\partial_q\left[\bigl(U'(q)+\nu g\bigr)\rho(q,t)\right]\nonumber\\
&&+ \nu \left[\rho (q-g,t) - \rho (q,t)\right]
\end{eqnarray}
(for completeness, we give a derivation in Appendix).

\subsection{A qualitative picture of large fluctuations}

We assume that the noise is weak and that initially the system is prepared well inside the potential well in Fig.~\ref{fig:metastable_potential}. Over time $\sim t_r=\la^{-1}$ the system will approach the vicinity of the potential minimum $q_a$ and will then fluctuate about $q_a$ for a long time $t\gg t_r$. Eventually there will happen a sequence of noise pulses that will cause the system to go over the barrier. In the time range $t_r\ll t\ll W^{-1}$ the probability distribution inside the potential well $\rho(q)$ is quasistationary. It is maximal near $q_a$ (but generally not exactly at $q_a$, see below).

Of primary interest to us is the tail of the distribution, which is formed by large fluctuations to states far from the equilibrium position, $(q-q_a)/g\gg 1$. Such fluctuations are rare and typically last for time $\sim t_r$, which is the characteristic time of the system dynamics. Because a single noise pulse shifts the system by $g$, the number $n$ of noise pulses required for reaching a point $q$ inside the potential well in time $t_f\sim t_r$ is determined by expression
\begin{equation}
\label{eq:number_of_pulses}
n-\nu t_f \sim(q-q_a)/g.
\end{equation}
Here, $\nu t_f$ is the average number of pulses in time $t_f$; the corresponding bias has been incorporated into the potential $U(q)$, cf. Eq.~(\ref{eq:Langevin}); the difference $n-\nu t_f$ characterizes the deviation of the noise from the average. In a rare fluctuation $|n-\nu t_f|\gg 1$, in agreement with the condition $(q-q_a)/g \gg 1$; cf. Ref.~\onlinecite{Billings2009} where the case $\nu t_r\lesssim 1$ was outlined.

The difference between the limits of small and large mean pulse rates is that $|n-\nu t_f|$ may be large or small compared to the average number of pulses $\nu t_f$. The case $n\gg \nu t_f$ corresponds to a comparatively low mean pulse rate. The case $\nu t_f\gg |n-\nu t_f|\gg 1$ corresponds to the large mean pulse rate limit, in which the noise is essentially Gaussian. This is clear from the expression for the probability to have $n$ pulses in time $t_f$,
\[P_n=(\nu t_f)^n\exp(-\nu t_f)/n!.\]
Indeed, the values of $n$ in the two limiting cases lie, respectively, on the non-Gaussian tail and in the Gaussian part of the distribution $P_n$.

For $n$ given by Eq.~(\ref{eq:number_of_pulses}), we have $\ln\rho(q)\sim \ln P_n$. In the limits of small and large pulse rate this gives, to leading order in $(q-q_a)/g$,
\begin{eqnarray}
\label{eq:estimates}
-\ln\rho(q)\sim \frac{(q-q_a)}{g}\ln\frac{(q-q_a)}{g\nu t_f},\quad \nu t_r\ll \frac{q-q_a}{g};\nonumber \\
-\ln\rho(q)\sim \frac{(q-q_a)^2}{2\nu g^2 t_f},\quad \nu t_r \gg \left|\frac{q-q_a}{g}\right|\gg 1
\end{eqnarray}
with $t_f\sim t_r$ (see Sec.~III~B). Equations (\ref{eq:estimates}) show that the tail of the distribution is qualitatively different depending on the mean pulse rate $\nu$.

Implicit in the estimate Eq.~(\ref{eq:number_of_pulses}) was the assumption that the noise is stronger than the regular force. With an appropriate $t_f$ this assumption gives the right answer not too far from $q_a$ where the potential $U(q)$ is parabolic. The parabolicity of $U(q)$ thus determines the range of applicability of Eq.~(\ref{eq:estimates}). Further away from $q_a$ Eq.~(\ref{eq:estimates}) still describes the leading-order term in $\ln\rho(q)$ for a low pulse rate, but becomes inapplicable for a high pulse rate.

Still further away from $q_a$, where $q$ is outside the potential well, for $t_r\ll t \ll W^{-1}$ the distribution $\rho(q)$ corresponds to a quasistationary probability current $j$, which is equal to the escape rate, $W=j$ \cite{Kramers1940}. One of our goals is to find $j$. This will be done by solving Eq.~(\ref{eq:Fokker_Planck}) separately in the three regions indicated in Fig.~\ref{fig:metastable_potential} and matching the solutions.

\section{Vicinity of the stable state}

We first consider region $I$ in Fig.~\ref{fig:metastable_potential}, where the system is close to the attractor, $|q-q_a|\ll |\qS-q_a|$. Here, the potential $U(q)$ can be expanded in $q-q_a$ keeping only the quadratic term, with $U'(q)\approx \la (q-q_a)$. This makes it possible to find an explicit solution of Eq.~(\ref{eq:Fokker_Planck}). For an overdamped particle in a parabolic potential driven by a unipolar Poisson noise with random pulse area, the probability distribution was found in Ref.~\cite{Baule2009}. In contrast, we consider the case of a constant pulse area, which is relevant for many physical sources of noise, including electron or photon noise; the method \cite{Baule2009} does not apply to this case and the distribution is different. The standard eikonal approximation often used for white-noise driven systems \cite{Freidlin_book} also does not apply.

We seek the solution of Eq.~(\ref{eq:Fokker_Planck}) with $U'=\la (q-q_a)$ in the form of a Fourier integral,
\begin{equation}
\label{eq:Fourier_near_stable}
\rho(q)=(2\pi)^{-1}\int_{-\infty}^{\infty}d\omega\exp\left[-i\omega (q-q_a)\right]\rho_a(\omega).
\end{equation}
Substituting Eq.~(\ref{eq:Fourier_near_stable}) into Eq.~(\ref{eq:Fokker_Planck}), we obtain a linear differential equation for $\rho_a(\omega)$, with solution
\begin{eqnarray}
\label{eq:soln_near_stable}
\rho_a(\omega)=\exp\left[\nu\la^{-1} \int\nolimits_0^{\omega}d\omega'\frac{e^{ig\omega'}-ig\omega' -1}{\omega'}\right].
\end{eqnarray}
Equation (\ref{eq:soln_near_stable}) gives $\rho_a(\omega)$ in terms of the exponential integral \cite{Abramowitz1972}. A constant factor in $\rho_a$ is chosen so that to satisfy the normalization condition $\int\nolimits_{-\infty}^{\infty}dq \rho(q)=1$.

The maximum of $\rho(q)$ is located close to the equilibrium position $q_a$. The shape of $\rho(q)$ near the maximum strongly depends on the parameter $\nu t_r\equiv \nu/\la$. For a high rate of noise pulses, $\nu/\la\gg 1$, one can expand the exponent in Eq.~(\ref{eq:soln_near_stable}) to the quadratic term in $\omega$, with the result
\begin{eqnarray}
\label{eq:gauss_near_stable}
&&\rho(q)\approx (\la/2\pi D)^{1/2}\exp\left[-\la(q-q_a)^2/2D\right], \nonumber\\
&&D=\nu g^2/2, \qquad |U'(q)/\nu g|\ll 1.\qquad
%\nu g^2\ll \la(\qS-q_a)^2.
\end{eqnarray}
Equation (\ref{eq:gauss_near_stable}) is the familiar Gaussian distribution near the minimum of a potential well for a particle driven by white Gaussian noise of intensity $\nu g^2/2$. The inequality in Eq.~(\ref{eq:gauss_near_stable}) is the condition that $|\omega g|\ll 1$ for the values of $\omega$ that give the main contribution to Eq.~(\ref{eq:Fourier_near_stable}); it is necessary for the expansion in $\omega$ to work.

The shape of the distribution near the maximum is qualitatively different in the opposite
limit of small mean pulse rate, $\nu\la^{-1}\ll (\qS-q_a)/g$, where we use $\qS-q_a$ as a typical distance on which the potential $U(q)$ becomes essentially nonparabolic. From Eq.~(\ref{eq:soln_near_stable}), for  $(q-q_a + g\nu \la^{-1})/g< 0$ the integrand in Eq.~(\ref{eq:Fourier_near_stable}) has no singularities for Im~$g\omega > 0$ and exponentially decays for Im~$g\omega\to \infty$. Therefore, by closing the integration contour over $\omega$ in the appropriate halfplane, we obtain $\rho(q)=0$ for $(q-q_a + g\nu \la^{-1})/g< 0$. We note that $q_a - g\nu\la^{-1}$ is still in the region where $U(q)$ is parabolic for small mean pulse rate; this is the equilibrium position of the ``bare" potential $U_0(q)$, i.e., the potential without the noise-induced bias, $U_0(q)=U(q)+g\nu q$. In contrast, in the Gaussian noise limit, Eq.~(\ref{eq:gauss_near_stable}), this point is generally beyond the range where $U(q)$ is parabolic.

The singular behavior of $\rho(q)$ for small to moderate $\nu/\la$ is easy to understand: unipolar noise pulses push the system only in the direction of positive $[q-q_a+ g\nu\la^{-1}]/g$. If the system has no inertia, its quasistationary distribution should indeed be zero for $(q-q_a + g\nu \la^{-1})/g< 0$ \cite{Billings2008,Baule2009}.
Because $|\rho_a(\omega)|$ decays as $|\omega|^{-\nu/\la}$ for large $\omega$, if $\nu<\la$ distribution $\rho(q)$ displays a power-law divergence,
\[\rho(q)\propto [(q- q_a + g\nu\la^{-1})/g]^{\nu\la^{-1}-1}\]
for $(q- q_a + g\nu\la^{-1})/g\to +0$, as is also the case for exponentially distributed heights of noise pulses \cite{Baule2009}. This is very different from the smooth Gaussian peak at the distribution maximum for large $\nu/\la$, cf. Eq.~(\ref{eq:gauss_near_stable}).

\subsection{Distribution tail in the harmonic region of the potential}

Of significant interest for us is the tail of the distribution, where $\rho(q)\ll 1$. It lies for $(q-q_a)/g\gg 1$ and is still in the harmonic region of $U(q)$ provided the typical width of the distribution peak is small compared to the typical distance $\qS-q_a$ on which $U(q)$ becomes nonparabolic. The distribution tail can be obtained from Eqs.~(\ref{eq:Fourier_near_stable}) and (\ref{eq:soln_near_stable}) for arbitrary $\nu/\la$. On the tail, integration over $\omega$ can be done by the method of steepest descent. The extremum of the integrand is reached for $\omega$ lying on the imaginary axis, $\omega=-ip$ with $p \equiv p(q)$ given by equation
\begin{eqnarray}
\label{eq:hamiltonian}
H(q,p)=0,\qquad H= \nu\left(e^{pg}-pg -1\right)-pU'(q).
\end{eqnarray}
In the range considered in this Section $U'(q)=\la (q-q_a)$, but the form in which Eq.~(\ref{eq:hamiltonian}) is written allows using the expression for $H(q,p)$ where $U'(q)$ is nonlinear in $q-q_a$, see below. The result of integration over $\omega$ can be put into the form
\begin{eqnarray}
\label{eq:explicit_near_attractor}
&&\rho(q)=\left(\la p/2\pi \partial_pH\right)^{1/2}\exp[-s(q)],\nonumber\\
&&s(q)=\int\nolimits_{q_a}^q dq'p(q').
\end{eqnarray}
Function $-s(q)$ is the Legendre transform of the exponent of $\rho_a(\omega)$ for $\omega=-ip(q)$. The method of steepest descent applies if $s(q)\gg 1$. We note that $s(q)$ can be thought of as an action of an auxiliary Hamiltonian system with coordinate $q$, momentum $p$, and Hamiltonian $H(q,p)$. Equation (\ref{eq:hamiltonian}) gives the Hamilton-Jacobi equation for the auxiliary system, $H(q,\partial_qs)=0$.

\subsubsection{Applicability of the steepest descent method}

The method of steepest descent applies, i.e., it is sufficient to keep in the exponent of $\rho_a(\omega)$ in Eqs.~(\ref{eq:Fourier_near_stable}) and (\ref{eq:soln_near_stable}) only terms quadratic in $(\omega + ip)$ , if the terms of higher order in $(\omega+ip)$ are small. A term $(\omega+ip)^n$ enters the expansion of $\ln\rho_a(\omega)$ with coefficient $K_n/n!$,
\begin{equation}
\label{eq:lof_rho_derivatives}
K_n= \left[\partial_{\omega}^n\ln\rho_a(\omega)\right]_{\omega=-ip}.
\end{equation}
One can check that $|K_n/g^{n-2}K_2|$ increases with $pg$ monotonically from $2/n$ for $pg\to 0$ to $1$ for $pg\to\infty$. Using that $K_2=-\la^{-1}\partial_pH/p$ [this relation has been used in deriving Eq.~(\ref{eq:explicit_near_attractor})] and taking into account the condition $s(q)\gg 1$, one finds the applicability condition of Eq.~(\ref{eq:explicit_near_attractor})
\begin{equation}
\label{eq:weak_noise}
\left|gU''(q)/\partial_pH\right| \ll(pg)^{-1},\qquad pU'(q)\gg |U''(q)|,
%pg\gg \min\left[1,(\la/\nu)^{1/2}\right].
\end{equation}
with $p(q)$ given by equation $H(q,p)=0$. Formally, Eq.~(\ref{eq:weak_noise}) was obtained for $U''(q)=\la, U'(q)=\la(q-q_a)$, but the form in which it is written makes it applicable also outside the range of parabolicity of $U(q)$, see below.

\subsection{The limits of almost Gaussian and strongly non-Gaussian noise }

The explicit form of the distribution on the tail can be obtained in the cases of comparatively large and small noise pulse rate. The limit of Gaussian noise corresponds to $\nu/\la\gg (q-q_a)/g\gg 1$. We have from Eq.~(\ref{eq:hamiltonian}) $p\approx 2\la(q-q_a)/\nu g^2\ll 1/g$ and $\partial_pH\approx \nu g^2 p/2$. The condition of being on the distribution tail, Eq.~(\ref{eq:weak_noise}), is met for $\la(q-q_a)^2/\nu g^2\gg 1$. For such $q$, Eq.~(\ref{eq:explicit_near_attractor}) coincides with Eq.~(\ref{eq:gauss_near_stable}). The exponent of $\rho(q)$ coincides also with the estimate Eq.~(\ref{eq:estimates}) for large $\nu/\la$ provided the typical duration of a fluctuation to $q$ is set equal to $t_f=t_r/2$; however, this is essentially an artifact, the assumption of the noise being much stronger than the regular force in an optimal fluctuation does not apply in the Gaussian-noise limit, see below.

In the opposite limit, $\nu/\la\ll (q-q_a)/g$, and in particular for $\nu/\la\lesssim 1$, on the tail of the distribution, where $(q-q_a)/g\gg 1$, we have from Eq.~(\ref{eq:hamiltonian}) $\exp(pg)\gg 1$. Then from Eq.~(\ref{eq:explicit_near_attractor}) we obtain for $s(q)$ an expression in which the leading-order term coincides with the right-hand side of the first equation in Eq.~(\ref{eq:estimates}) if we set $t_f=et_r$. We note that the prefactor in the distribution $\rho(q)$, Eq.~(\ref{eq:explicit_near_attractor}), now explicitly depends on $q$. To the leading order it is $\sim [2\pi g(q-q_a)]^{-1/2}$. The first condition in Eq.~(\ref{eq:weak_noise}) reduces to $U'(q)/g\gg U''(q)$ and is satisfied for $(q-q_a)/g\gg 1$; clearly, the second condition in Eq.~(\ref{eq:weak_noise}) is then also satisfied.

\section{The region away from the stationary states}

We now consider region $II$ in Fig.~\ref{fig:metastable_potential}, where coordinate $q$ is far from the both stationary positions of noise-free motion, $|q_a-q|, |\qS-q|\gg |g|$. In the spirit of the WKB approximation, we seek the quasi-stationary probability distribution $\rho(q)$ in this range in the eikonal form,
\begin{equation}
\label{eq:eikonal:general}
\rho(q)=\exp[-S(q)], \qquad S(q)\gg 1.
\end{equation}
We substitute this expression into Eq.~(\ref{eq:Fokker_Planck}) for $\rho(q)$ and expand $S(q-g) \approx S(q)- gP +(1/2)g^2P'$, where $P=S'$. The expansion is justified if $S$ and $P$ vary smoothly on the distance $\sim g$, even though $\rho(q)$ does not. Respectively, we will assume that $Pg$ is not necessarily small; however, as will be shown later, $g^2|P'|\ll 1$.

We seek $P$ in the form $P\approx P^{(0)} + P^{(1)}$, with $|P^{(1)}|\ll |P^{(0)}|$, and respectively, $S\approx S^{(0)}+S^{(1)}$. To find the leading order term, $P^{(0)}$, we can disregard the term $\propto P'$ in the expansion of $S$; we can also disregard $U''\rho$ compared to $U'P^{(0)}\rho$. Then we obtain from Eq.~(\ref{eq:Fokker_Planck})
\begin{equation}
\label{eq:zeroth_order_P}
P^{(0)}= p\equiv p(q),\qquad S^{(0)}=s(q)\equiv\int\nolimits_{q_a}^qp(q)dq,
\end{equation}
where $p(q)$ is given by equation $H(q,p)=0$ with $H$ defined by Eq.~(\ref{eq:hamiltonian}); note that here we do not assume that $U'(q)$ is linear in $q$.

The term $P^{(1)}$ is given by a linear equation which follows from Eqs.~(\ref{eq:Fokker_Planck}) and (\ref{eq:eikonal:general}) if we keep linear terms in $P^{(1)},P^{(0)\prime}\equiv p^{\,\prime}$. Using the relation
\begin{equation}
\label{eq:p_derivative}
p^{\,\prime}\equiv dp/dq = pU''(q)/\partial_pH,
\end{equation}
which follows from Eqs.~(\ref{eq:hamiltonian}), one finds after some algebra
\begin{equation}
\label{eq:first_order_P}
P^{(1)}(q)= (2\partial_p H)^{-1}\left[\frac{d}{dq}\partial_pH - U''(q)\right].
\end{equation}

From Eqs.~(\ref{eq:eikonal:general})--(\ref{eq:first_order_P}) we obtain
\begin{equation}
\label{eq:explicit_eikonal}
\rho(q)\approx C\left(p/\partial_pH\right)^{1/2}e^{-s(q)},\qquad C=(\la/2\pi)^{1/2}.
\end{equation}
Here, $s(q)$ is given by Eqs.~(\ref{eq:hamiltonian}) and (\ref{eq:explicit_near_attractor}). The constant $C$ is chosen in such a way as to match the probability distribution (\ref{eq:explicit_near_attractor}) in the region where $U(q)$ is parabolic but $(q-q_a)/g\gg 1$.

\subsubsection{Applicability of the eikonal approximation}

The condition of applicability of the eikonal approximation is more complicated than in the simple and well-known cases of white Gaussian noise or reaction (birth-death) systems, cf. Refs. \onlinecite{Freidlin_book,Escudero2009}. This is because the system does not have one small parameter: even though $g/(\qS-q_a)$ is small, we may still have $\nu\la^{-1}\gg (\qS-q_a)/g$. We now show that Eq.~(\ref{eq:weak_noise}) provides sufficient applicability conditions in the whole range of interest.

It is immediately seen from Eq.~(\ref{eq:p_derivative}) that, where Eq.~(\ref{eq:weak_noise}) holds, we have $g^2|P^{(0)\prime}|\ll 1$, as assumed.  To check that $|P^{(1)}|$ is small and thus to justify the used expansion, one can rewrite Eq.~(\ref{eq:first_order_P}) as $P^{(1)}= pU''\Pi/2(\partial_pH)^2$ with $\Pi=\partial^2_pH-2p^{-1}\partial_pH$. It follows from condition $H(q,p)=0$ that the ratio
$ \Pi/g\partial_pH$ increases monotonically from $2/3$ for $0\leq pg \ll 1$ to $1$ for $pg\to\infty$. Therefore
\[P^{(1)}\sim pgU''/\partial_pH.\]
From this relation one can see that Eq.~(\ref{eq:weak_noise}) indeed leads to $|P^{(1)}|\ll |p|, 1/g$; in fact, one should separately consider the case $pg < 1$, but the overall analysis is straightforward.

\subsection{Limiting cases}
The explicit form of the distribution $\rho(q)$ is easy to find in the limit of high rate of noise pulses,  $\nu\gg U'(q)/g$, where the noise is perceived by the system as Gaussian. In this case from $H=0$ we find $p=2U'/\nu g^2$, and then from Eq.~(\ref{eq:explicit_eikonal})
\begin{eqnarray}
\label{eq:Gauss_distr_general}
&&\rho(q)=\left(\la/2\pi D\right)^{1/2} e^{-\left[U(q)-U(q_a)\right]/D}.
%&&D=\nu g^2/2.\nonumber
\end{eqnarray}
This is the Boltzmann distribution for a particle in a potential $U(q)$; the effective temperature $D=\nu g^2/2$ is equal to the noise intensity. In the range where $U(q)$ is parabolic, Eq.~(\ref{eq:Gauss_distr_general}) coincides with Eq.~(\ref{eq:gauss_near_stable}) obtained in a different way.

In the opposite limit of low to moderate rate of noise pulses, $\nu\ll U'(q)/g$, we have
\begin{eqnarray}
\label{eq:nonGauss_general}
\rho(q)\approx \left[\la/2\pi g U'(q)\right]^{1/2}\exp[-s(q)],
\end{eqnarray}
where $s(q)$ is given by Eq.~(\ref{eq:zeroth_order_P}) with $p\approx g^{-1}\left[\ln(U'/\nu g)+ \ln\ln(U'/\nu g)\right]$ \cite{Billings2009}. In the range $U'=\la(q-q_a)$ Eq.~(\ref{eq:nonGauss_general}) goes over into the result obtained in Sec.~III in a different way.

For low pulse rate the value of $s(q)$ is close to what follows from the simple expression Eq.~(\ref{eq:estimates}) if one estimates the duration of the fluctuation in Eq.~(\ref{eq:estimates}) as $t_f\sim (q-q_a)/U'(q)$, which is reasonable (the value of $s$ depends on $t_f$ logarithmically). We remind that Eq.~(\ref{eq:estimates}) was obtained assuming that the optimal train of the noise pulses that bring the system to a remote state $q$ gives a much stronger force than the regular force. As seen from Eq.~(\ref{eq:Gauss_distr_general}), this assumption does not apply to the case where the mean pulse rate is high and the noise is perceived as Gaussian. For Gaussian noise, the optimal force is known \cite{Luchinsky1998}, for white noise it is equal to twice the regular force.

\section{Vicinity of the local potential maximum}

Near the local potential maximum, region $III$ in Fig.~\ref{fig:metastable_potential}, the potential is parabolic, $U(q)\approx U(\qS)-\lS (q-\qS)^2/2$. The width of the region where the parabolic approximation applies largely exceeds $|g|$. In a part of this region deep inside the potential well the distribution is described by Eq.~(\ref{eq:explicit_eikonal}). For a small mean rate of noise pulses, $\nu/\lS\lesssim 1$, the distribution has an important qualitative feature, the square-root divergence of the prefactor for $q$ approaching the stationary state $\qS$, which is seen from Eq.~(\ref{eq:nonGauss_general}). This divergence is smeared out at a distance $\sim |g|$ from $\qS$, see below, but it makes it impossible to use the Kramers method of finding the distribution near the potential maximum \cite{Kramers1940}, since the method substantially relies on the smoothness of the distribution on a scale that largely exceeds $g$.

Sufficiently far outside the well, on the other hand, the situation is similar to that discussed by Kramers. The effect of weak noise can be disregarded here. The distribution describes a coordinate-independent quasistationary probability current $j$ from the potential well. Close to $\qS$, but for $(q-\qS)/g\gg 1$
\begin{equation}
\label{eq:escape_rate_definition}
W=j\approx \lS (q-\qS)\rho(q).
\end{equation}
To simplify notations, we assume here that $g>0$; the generalization to the case $g<0$ is straightforward. Equation~(\ref{eq:escape_rate_definition}) follows from the kinetic equation, Eq.~(\ref{eq:Fokker_Planck}), upon integration over $q$ from $-\infty$ to a given $q>\qS$, with account taken of the smoothness of $\rho(q)$ on a distance $\sim g$. The disregarded correction is $\propto \nu g^2/\lS(q-\qS)^2 \ll 1$; note that we do not assume that $\nu g/\lS(q-\qS)$ is small, the terms linear in $g$ drop out from the expression for $j$.

In order to allow for the singular behavior of $\rho(q)$ for $q$ approaching $\qS$ from inside the well, we seek the solution of Eq.~(\ref{eq:Fokker_Planck}) with $U'=-\lS(q-\qS)$ in the form of a Laplace transform, as done previously for periodically modulated systems driven by white noise \cite{Smelyanskiy1999,Dykman2005a},
\begin{equation}
\label{eq:near_saddle_general}
\rho(q)=\int\nolimits_0^{\infty}dk e^{-k(q-\qS)/g}\rho_{\cal S}(k).
\end{equation}
Substituting this ansatz into Eq.~(\ref{eq:Fokker_Planck}) and solving the resulting first-order equation for $\rho_{\cal S}(k)$, we find
\begin{equation}
\label{eq:near_saddle_Laplace}
\rho_{\cal S}(k)=C_{\cal S}\exp\left[-\frac{\nu}{\lS}\int_0^kdk'\left(e^{k'}-k'-1\right)/k'\right].
\end{equation}

Deep inside the well, but still in the region where $U(q)$ is parabolic, the integral over $k$ in Eq.~(\ref{eq:near_saddle_general}) can be calculated by the steepest descent method. The extremum with respect to $k$ lies for $k=pg$, where $p\equiv p(q)$ is determined by Eq.~(\ref{eq:hamiltonian}) with $U'=-\lS(q-\qS)$. The result of the integration over $k$ is
\begin{eqnarray}
\label{eq:explicit_near_saddle_inside}
\rho(q)\approx C_{\cal S}\left[\frac{2\pi\lS g^2 p}{\partial_pH}\right]^{1/2} \exp\left[-\int\nolimits_{\qS}^qp(q)dq\right].
\end{eqnarray}
We note that $\rho(q)$ increases with $q$ moving inside the well, i.e., with increasing $(\qS-q)/g$.

An analysis completely analogous to that in Sec.~III~A~1 shows that the steepest descent method applies for sufficiently large $(\qS-q)/g$ so that there hold inequalities (\ref{eq:weak_noise}). For $(\qS-q)/g\gg 1$ and for low mean pulse rate $\nu/\lS \lesssim 1$, we have $\exp(pg)\gg 1$, which means that the distribution $\rho(q)$ changes by a large factor when $q$ changes by $g$. The smoothening of the distribution occurs in the narrow range $|(\qS-q)/g| \lesssim 1$; the steepest descent method does not work in this range.

Equations (\ref{eq:explicit_eikonal}) and (\ref{eq:explicit_near_saddle_inside}) match in a broad range of $q$, provided we set
\begin{equation}
\label{eq:near_saddle_constant}
C_{\cal S}= \frac{1}{2\pi |g|}(\la/\lS)^{1/2}e^{-Q},\qquad Q= \int_{q_a}^{\qS}p(q)dq.
\end{equation}

On the other hand, outside the potential well for $(q-\qS)/g\gg 1$, the major contribution to the integral over $k$ in Eq.~(\ref{eq:near_saddle_Laplace}) comes from small $k\lesssim g/(q-\qS)\ll 1$. For such $k$, $\rho_{\cal S}(k)\approx C_{\cal S}$, and from Eq.~(\ref{eq:near_saddle_general}) $\rho(q)\approx C_{\cal S}g/(q-\qS)$. Using the value of $C_{\cal S}$ from Eq.~(\ref{eq:near_saddle_constant}), one obtains from Eq.~(\ref{eq:escape_rate_definition}) an explicit expression for the escape rate,
\begin{equation}
\label{eq:explicit_escape_rate}
W=\frac{1}{2\pi}(\la\lS)^{1/2}\exp(-Q).
\end{equation}
This is the central result of the paper.

\begin{figure}[h]
\begin{center}
\includegraphics[width=2.0in]{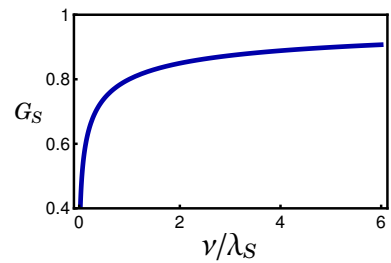}
\end{center}
\caption{The ratio of the scaled probability density at the saddle point to the escape rate $G_{\cal S}$, Eq.~(\ref{eq:G_S}). }
\label{fig:saddle_ratio}
\end{figure}

It is instructive to look at the ratio of the probability distribution to find the particle at the local maximum of the potential $\qS$ and the escape probability $W$. We normalize $\rho(q)$ by the characteristic diffusion length in the Gaussian-noise limit $(\nu g^2/\pi\lS)^{1/2}$ and $W$ by $\lS$ and introduce function $G_{\cal S}$,
\begin{eqnarray}
\label {eq:G_S}
G_{\cal S}=\left(\nu\lS g^2/\pi\right)^{1/2}\rho(\qS)/W.
\end{eqnarray}
For the chosen normalization, $G_S$ approaches 1 in the limit of large $\nu/\lS$. On the other hand, for small $\nu/\lS$ we have from Eqs.~(\ref{eq:near_saddle_general}), (\ref{eq:near_saddle_Laplace}), (\ref{eq:near_saddle_constant}), (\ref{eq:explicit_escape_rate}) $G_{\cal S}\propto (\nu/\lS)^{1/2}|\ln(\nu/\lS)|$. The overall dependence of $G_{\cal S}$ on $\nu/\lS$ is shown in Fig.~\ref{fig:saddle_ratio}. It is seen that $G_{\cal S}$ approaches the Gaussian noise limit comparatively slowly, and that it significantly differs from the Gaussian-noise result for $\nu/\lS\lesssim 1$.

\section{Discussion}

We have studied the probability distribution and the rate of escape from a potential well of an overdamped particle driven by Poisson noise. The results cover a broad range of the average rate of noise pulses $\nu$, from small rates, where $\nu$ is comparable to the reciprocal system relaxation time $t_r^{-1}$, to high rates, $\nu\gg t_r^{-1}$. Noise-induced fluctuations are very different in these limits. For $\nu t_r\lesssim 1$ the state of the system changes between successive pulses, if they come at the average rate, and thus the discreteness of the pulses is essential. For $\nu t_r\gg 1$ where, on average, many pulses occur within the relaxation time, the system may perceive the noise as effectively Gaussian.

The noise is assumed weak, so that the rate of noise-induced escape $W$ is small compared to $t_r^{-1}$. Escape results from a large rare fluctuation, which is an unlikely sequence of individual noise pulses, with the overall duration of the fluctuation $\sim t_r$.
The condition $Wt_r\ll 1$ requires that the noise pulse area $g$ be small compared to the distance between the stationary states, $(\qS-q_a)/g \gg 1$, which means that many noise pulses are needed for escape. However, the developed theory is not just an asymptotic theory for $g\to 0$. This makes it significantly different from the theories of escape due to Gaussian noise where the noise intensity is the smallest parameter, cf. Refs.~\onlinecite{Freidlin_book,Luchinsky1998} or due to reaction randomness in reaction systems, where the reciprocal number of particles is the small parameter, cf. Refs. \onlinecite{Dykman1995c,Escudero2009}.

In our case, the Gaussian noise limit corresponds to $\nu t_r \gg (\qS-q_a)/g$, which is incompatible with the limit $g\to 0$. Formally, Poisson noise is characterized by two parameters, $\nu$ and $g$, and it is the interrelation between these parameters and the system parameters that leads to the rich pattern of fluctuations.

In the Gaussian noise limit, our results coincide with the well known results for this case. The distribution is of the Boltzmann form with temperature given by the noise intensity $D=\nu g^2/2$. The escape rate is described by the Kramers theory \cite{Kramers1940}. In the Gaussian case the time interval between the noise pulses in the most probable fluctuation leading to escape varies as the fluctuation progresses, keeping the velocity of the system equal to $U'(q)$ not too close to the stationary states.

For $\nu t_r\ll (\qS-q_a)/g$, on the other hand, the probability distribution differs qualitatively from the Boltzmann distribution, it is singular near the maximum. Away from the maximum, along with a non-Boltzmann exponential factor it contains a coordinate-dependent prefactor. The latter scales as an inverse square root of the potential gradient $U'(q)$, Eq.~(\ref{eq:nonGauss_general}). Here, the time interval between the noise pulses in the most probable fluctuation leading to escape is such that the velocity largely exceeds $U'(q)$ far from the attractor and the saddle point.

The explicit expression for the probability distribution allows one to see the evolution of the distribution depending on the parameter $\nu t_r$, i.e., with the noise varying from strongly non-Gaussian to effectively Gaussian. We note that the standard WKB-type approximation does not apply near the maximum of the distribution for $\nu t_r\ll (\qS-q_a)/g$.

The rate of Poisson-noise induced escape has a simple form, Eq.~(\ref{eq:explicit_escape_rate}). It contains
an exponential factor $\exp(-Q)$ and the prefactor $(\la \lS)^{1/2}/2\pi$. The exponent $Q$ is given by Eqs.~(\ref{eq:hamiltonian}) and (\ref{eq:near_saddle_constant}). In contrast to escape due to white Gaussian noise, for Poisson noise $Q$ is not determined by the height of the potential barrier, but depends on the actual shape of the potential well. For $\nu t_r\ll (\qS-q_a)/g$ it is particularly sensitive to the distance between the maximum and minimum of the potential \cite{Billings2008,Billings2009}. Also, $Q$ depends not on the noise intensity $\nu g^2$, but separately on $\nu$ and $g$.

The prefactor in the escape rate is independent of the noise parameters and is determined only by the shape of the potential near its local minimum and maximum. Remarkably, it has the same form as the celebrated Kramers expression for the case of an overdamped system driven by white Gaussian noise \cite{Kramers1940}. This is unexpected, given that the shape of the distribution is generally qualitatively different from the Boltzmann distribution, and the analysis used to obtain the rate is very different from the Kramers analysis.

I am grateful to A. Kamenev, M. Khasin, and B. Meerson for useful discussions. This research was supported in part by NSF grant CMMI-0900666.

\appendix

\section{Kinetic equation for a Poisson-noise driven system}
\label{sec:kinequation}

We write the probability density as $\rho(q,t)=\langle\delta[q-q(t)]\rangle$, where the averaging is performed over realizations of noise $f_P(t)$, and $q(t)$ is given by the Langevin equation, Eq.~(\ref{eq:Langevin}). Then
\begin{widetext}
\begin{eqnarray}
\label{eq:increment_rho}
&&\rho(q,t+\Delta t)=\frac{1}{2\pi}\int dk\bigl\langle\exp\left\{ik\left[q-q(t)\right]+ik\int\nolimits_t^{t+\Delta t}dt_1 U'\bigl(q(t_1)\bigr)+ik\nu g\Delta t\right\}I_k[f_P]\bigr\rangle,\\ &&I_k[f_P]=\exp\left[-ik\int\nolimits_t^{t+\Delta t}dt_1f_P(t_1)\right].\nonumber
\end{eqnarray}

For small $\Delta t$, we can replace $U'\bigl(q(t_1)\bigr)$ with $U'\bigl(q(t)\bigr)$.
Since $q(t)$ is independent of $f_P(t_1)$ for $t_1>t$, the averaging of $I_k[f_P]$ can be done separately from the first exponential in Eq.~(\ref{eq:increment_rho}). Using the explicit form of the characteristic functional for a Poisson noise \cite{FeynmanQM}, we obtain
\begin{eqnarray}
\label{eq:character_fnctnl}
\langle I_k[f_P]\rangle= \exp\left[-\nu\,\Delta t\left(1-e^{-ikg}\right)\right]\approx 1-\nu\,\Delta t\left(1-e^{-ikg}\right).
\end{eqnarray}
We note also that
\begin{eqnarray}
\label{eq:increment_first_order}
%&&
\frac{1}{2\pi}\bigl\langle\int dk\left\{ik\Delta t\left[ U'\bigl(q(t)\bigr)+\nu g\right]\right\}\exp\left\{ik\left[q-q(t)\right]\right\}\bigr\rangle \approx \Delta t\partial_q\left\{\left[U'(q)+\nu g\right]\rho(q,t)\right\}.
%, \nonumber\\
%&&\frac{1}{2\pi}\bigl\langle\int dk\exp\left\{ik\left[q-q(t)\right]\right\}e^{-ikg}\bigr\rangle =\rho(q-g,t).
\end{eqnarray}
In the limit $\Delta t\to 0$, Eqs.~(\ref{eq:increment_rho})-- (\ref{eq:increment_first_order}) immediately give kinetic equation (\ref{eq:Fokker_Planck}).

\end{widetext}

%\bibliographystyle{apsrev}
%\bibliography{c:/Aaa/Bibtex/md10}

\end{document}